# A Data Colocation Grid Framework for Big Data Medical Image Processing – Backend Design

Shunxing Bao*[a], Yuankai Huo[b], Prasanna Parvathaneni[b], Andrew J. Plassard[a], Camilo Bermudez[c], Yuang Yao[a]
Ilwoo Lyu[a], Aniruddha Gokhale[a], Bennett A. Landman[a,b,c]

[a] Computer Science, Vanderbilt University, Nashville, TN, USA 37235
[b] Electrical Engineering, Vanderbilt University, Nashville, TN, USA 37235
[c] Biomedical Engineering, Vanderbilt University, Nashville, TN, USA 37235

## ABSTRACT

When processing large medical imaging studies, adopting high performance grid computing resources rapidly becomes important. We recently presented a "medical image processing-as-a-service" grid framework that offers promise in utilizing the Apache Hadoop ecosystem and HBase for data colocation by moving computation close to medical image storage. However, the framework has not yet proven to be easy to use in a heterogeneous hardware environment. Furthermore, the system has not yet validated when considering variety of multi-level analysis in medical imaging. Our target design criteria are (1) improving the framework's performance in a heterogeneous cluster, (2) performing population based summary statistics on large datasets, and (3) introducing a table design scheme for rapid NoSQL query. In this paper, we present a heuristic backend interface application program interface (API) design for Hadoop & HBase for Medical Image Processing (HadoopBase-MIP). The API includes: Upload, Retrieve, Remove, Load balancer (for heterogeneous cluster) and MapReduce templates. A dataset summary statistic model is discussed and implemented by MapReduce paradigm. We introduce a HBase table scheme for fast data query to better utilize the MapReduce model. Briefly, 5153 T1 images were retrieved from a university secure, shared web database and used to empirically access an in-house grid with 224 heterogeneous CPU cores. Three empirical experiments results are presented and discussed: (1) load balancer wall-time improvement of 1.5-fold compared with a framework with built-in data allocation strategy, (2) a summary statistic model is empirically verified on grid framework and is compared with the cluster when deployed with a standard Sun Grid Engine (SGE), which reduces 8-fold of wall clock time and 14-fold of resource time, and (3) the proposed HBase table scheme improves MapReduce computation with 7 fold reduction of wall time compare with a naïve scheme when datasets are relative small. The source code and interfaces have been made publicly available.

## 1. INTRODUCTION

When processing large medical imaging studies, adopting high performance grid computing resources rapidly becomes important. An inexpensive solution is to locate the data on the computational nodes to avoid the problem of saturating the network by copying data. For example, the Apache Hadoop Ecosystem provides an extensive suite of tools to co-locate storage and computation [1-3]. Hadoop is still not widely being integrated into medical image processing (MIP), although it has shown great success in online commerce [4-6], social media [7, 8], and video streaming [9-11]. Several recent medical image processing (MIP) approaches have aimed to take advantage of this big data architecture for specific use cases with MapReduce and distributed systems [12-15]. We recently presented a "medical image processing-as-a-service" grid framework, Hadoop & HBase for Medical Image Processing (HadoopBase-MIP), which integrates the Hadoop and HBase (a database built upon Hadoop) for data colocation by moving computation close to medical image storage [16-18]. This system is a general framework for MIP (e.g., structured data retrieval, access to locally installed binary executables/system resources, structured data storage) without commingling idiosyncratic issues related to MIP. However, the system has not yet proven ease to use, and faces several key challenges for wide deployment as illustrated in Figure 1. Specifically,

    1) Heterogeneous hardware: Hadoop/HBase uses an approximately balanced data allocation strategy by default. In [18], we observe that the throughput of a cluster is low especially when it combines with different types and / or different number of cores per machine. Thus, performance aware data collocation models are needed.

    2) Large dataset analysis: When large summary statistical analyses are requested, huge volumes of data can saturate memory of one machine. Thus, one needs to split a dataset into small chunks. Most chunks would be sent to where their data is located and run in parallel on different machines, but few chunks would need to move data via network. This process is called Map phase in MapReduce computation paradigm [1, 2]. Once all intermediate results of all chunks are collected on one machine, they are further processed and final result is generated, this is depicted

as Reduce phase [1, 2]. However, as presented in Figure 1 (B), integrated processing of large datasets introduces dependencies with the number of jobs (which is based on chunk size) and a way to optimize chunk size is needed.

3) Rapid NoSQL query: HBase is a column based NoSQL database, namely all data that are in same column are stored together [2, 19, 20]. Each row of table is a multi-record that are based on total number of columns are defined with a unique name as rowkey. If we store all medical images with other data like index, age and sex into same column, then when we perform a query, linear search with image traversal is required, which rapidly decreases the speed of search. On the other hand, for HBase MapReduce, each single map needs to set a start/stop row and a typical column. When doing subset datasets analysis, data is scattered in HBase, and not all record between start/ stop row are needed. It is important to structure data in a manner to skip unnecessary row and image traversal.

To resolve these challenges, we present a novel interface application program interface (API) that is based on Hadoop and HBase. We make following contributions: 1) offer better data allocation in a heterogeneous cluster for faster processing through an off-line load balancer; 2) describe an optimal criterion to find the splitting chunk size of large dataset; 3) enable fast data query and boot up the of MapReduce performance by using a HBase table scheme.

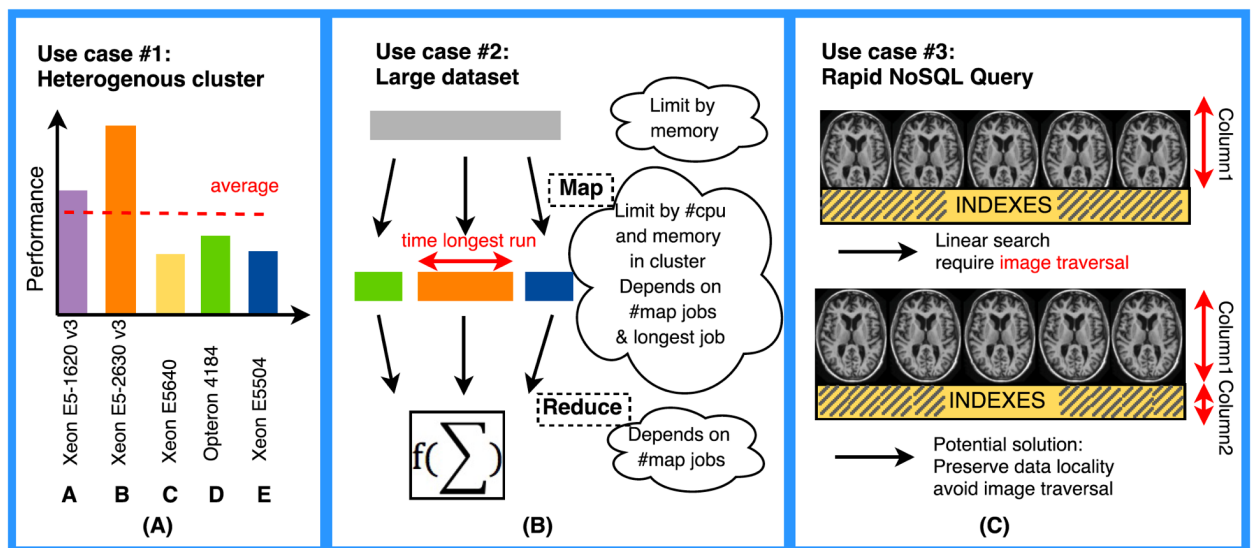

Figure 1. Use cases for three main challenges. (A) If a traditional cluster model is used, average throughput would be seen (red dash), which would leave some machines starved (e.g. A, B), while others overloaded (e.g. C, D and E). Hence, a traditional approach will degrade the overall execution time due to those overloaded/starved cores. (B) The time to run a large dataset depends on total number of jobs and the longest map job to take. The total number of jobs should be neither too large or too small. (C) HBase is not designed for storing image data given variability of size and volume of medical imaging studies. If information like age / sex / genetics are stored in same column with image data, image traversal is unavoidable, which degrades the search efficiency.

## 2. METHODS

### 2.1 HadoopBase-MIP system interface

A simplified hierarchical data storage model of HBase is: Table -> Column family -> Column qualifier -> data. An HBase table can contain many columns, and a column is usually denoted as column family. We can define many column qualifiers to specify a column family. Data is stored within qualifiers. Each row of the table has a unique rowkey. To retrieve or delete a set of data, users need to specify *(table name, rowkey, column family: column qualifer)*. Alternatively, one can retrieve a sequences of values by assigning a table rowkey range with relative column family and qualifier. A table is split into different regions. If any size of regions exceeds a pre-set threshold, then regions would be split into two new regions. This split process is activated through region split policy. Regions will be dispatched to different machines of cluster for balancing issue.

Figure 2 presents our system interface overview and how they may help tackle three challenges in section 1, which includes Upload, Retrieve, Load Balancer, MapReduce Template, Delete and Monitor. All operations are command-line based software tool with different parameters. Table 1 presents more detailed descriptions.

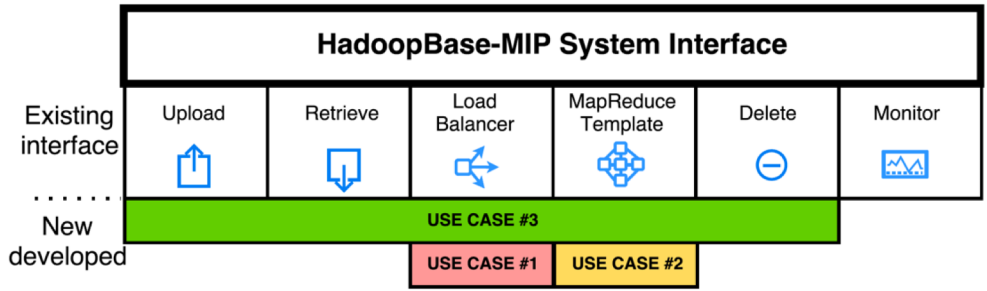

Figure 2. HadoopBase-MIP system interface overview. Except cluster monitoring, all operations are extended.

Table 1. HadoopBase-MIP system interface description

| Operation | Interface Parameter Description |
|---|---|
| **Upload** | - **Table name**: will create a table if it does not exist.<br>- **A text file path** for a file contains groups of <column family, qualifier> - to create / alter a table scheme.<br>- **A text file path** for a file contains all images' tuple of <system file path, file unique name, column family, qualifier>; File unique name will be used as rowkey.<br>- **Overwrite** – Boolean value. It helps update images or avoid uploading duplicate data.<br>- **Region split policy** : default policy, hierarchical policy [2]<br>- **Pre-split**: Boolean value. It is only valid when creating a new table.<br>- **A text file path** for a file contains all rowkeys for pre-split a table. |
| **Retrieve** | - **Table name**<br>- **Rowkey**: set this value if retrieval is image based<br>- **Start rowKey** and / or **stop row key**: set them for a retrieval range, if both keys are empty, then retrieval is whole table column based.<br>- **Column family**<br>- **Column qualifier**<br>- **A text file path** for a file contains all data retrieval destination path.<br>- **A text file path** for a file contains all row keys need to skip to retrieve. |
| **Delete** | All options are same with Retrieve except a text file path for a file contains all data retrieval destination path. |
| **Monitor** | Hadoop built in job monitoring tool. |
| **MapReduce Template** | - **Table name**: two names, one for source table to read data, another one for target table to write back result.<br>- **Column family**: three values, one for data query (will introduce more in section 2.3), one for image data retrieval, the last one for target table.<br>- **Column qualifier**: three correspondence value with column family.<br>- **A text file path** for a file contains all start / stop row key pair, each pair is input for a map job.<br>- **A text file path** for a file contains all row keys need to skip to retrieve.<br>- **Analysis level** – three options, image-based [18], dataset based [17] and large dataset based (will introduce more in section 2.2 & 2.3). |
| **Load Balancer** | HBase default built in balancer is to balance the total number of region on each server. Our proposed load balancer is offline greedy allocation. First it finds all regions and images on each serve; second, moving images based on region; finally, the data allocation ratio of each machine meets the ratio of (total number of CPU * Million instructions per seconds (MIPS)) per nodes. MIPS is calculated by Linux perf and is varied based on different types of CPU.<br>- **Table name**<br>- **Column family**<br>- **Column qualifier** |

## 2.2 MapReduce model design and implementation for large datasets

As an exemplar use case, we consider processing of T1-weigthed brain MRI images and construction of population specific averaged brain image. Creating a specific template for a group of studies can help improve finding cortical folding pattern and reduce misclassification of brain tissue[21-23]. In this subsection, we will present a MapReduce computation model in HadoopBase-MIP to average a large dataset. Figure 3 shows a full model pipeline to process a large dataset using MapReduce analysis. We first present a model for the pipeline considering two important factors: wall clock time (what the user directly experiences) and resource time (time elapsed time on each node when a process starts across all nodes). We summarize the parameters and helper functions that affect both wall-clock time and resource time theoretical model for finding optimizing map task chunk size in Table 2.

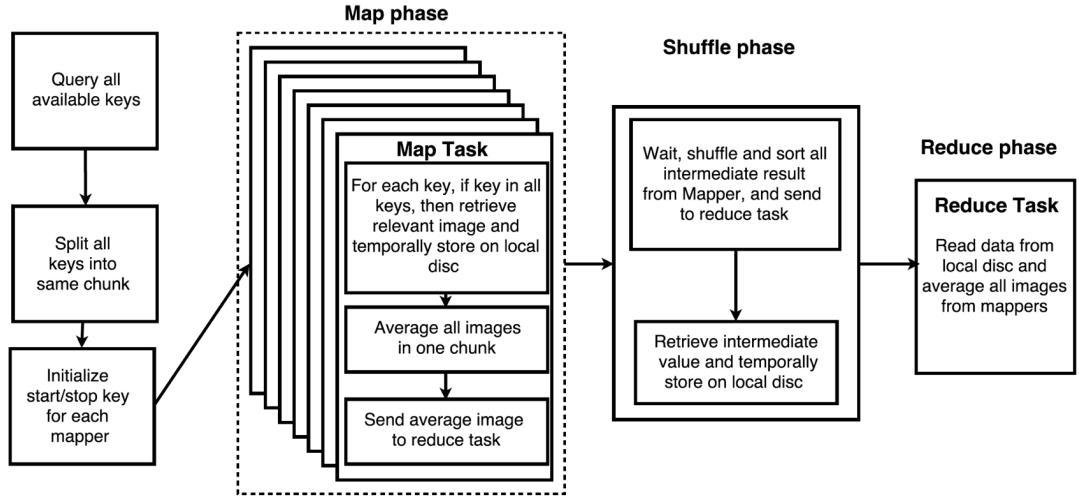

Figure 3. MapReduce model implementation for constructing population specific brain MRI templates.

Table 2. Model of wall time and resource time for average analysis parameters definition

| Definition | Description |
|---|---|
| $\#img$ | The total number of images that are needed to be averaged. |
| $\#job$ | The total number of map jobs that are split from large datasets. |
| η | The total number of images per map task, which is the chunk size. It helps find the total number of map jobs: $\#job = \#img/η$. **This is the variable that we are trying to find an optimizing value.** We assume there is no local weighted concern in split map tasks, it means all map tasks share the same value of η. |
| $SizeBig$ | Maximum input file size of datasets, we use it for worst case scenario estimation. |
| $SizeSmall$ | Minimum input file size of datasets, we use it for upper and lower bound of η. |
| $SizeGen$ | Maximum output file size that is generated by image averaging software. |
| $Bandwidth$ | The bandwidth of cluster. |
| $VdiscR, VdiscW$ | Data read / write speed of local hard drive. |
| $\#region$ | The total number of regions of a table in cluster. |
| $mem$ | The total memory of one machine. We presume all machines have same amount of memory. |
| $core$ | The total number of CPU cores of the cluster. |
| α | When map tasks generate intermediate results, part of them are stored in network buffer, and others are flushed into local temporarily and transfer to reduce task's shuffle phase later. α is unbuffered ratio of map tasks' results due to limit of heap size and cannot be held in memory. |
| β | β is an experimental empirical parameter to represent the ratio of rack-local map task for Hadoop scenario, namely the data is loaded/stored via network. We empirically get its value with 0.9. |
| $discR(x)$ $discW(x)$ | x is the size of a file. The function is used to calculate the time to read / write a file from / to local disc, namely $discR(x) = x/VdiscR$; $discW(x) = x/VdiscW$ |
| $bdw(x)$ | x is the size of a file. The function is used to calculate the time to transfer through network, namely $bdw(x) = x/Bandwidth$ |

| $avgANTS(\eta)$ | x is the size of a file. We use ANTS AverageImages [24] to empirically test average summary statistics analysis. We also do several profiling experiments to model this image processing. However, it is hard to conclude a concrete model for ANTS average to match all profiling results. We can prove the total number of file sizes that are needed to be averaged ($\eta \cdot x$) grows much slower than chunk size η itself. And the best solution is doing all average on 1 CPU. We found a worse case of $avgANTS(\eta) = 0.4\eta + 5$ to illustrate this process. |
|---|---|

In Map phase, $\eta$ is lower limited by total number of CPU cores since we aim to use one round to execute all map tasks. The upper bound is a confined by machine's memory. In Reduce phase, $\#job$ depends on $\eta$ and further limited by machine's memory. In summary, the range of valid $\eta$ is defined as $\eta \in \left[ \max\left( \frac{\#img \cdot SizeSmall}{mem}, \frac{\#img}{core} \right), \frac{mem}{SizeBig} \right]$.

Equation (1) is an overview wall-clock time model.

$$WT_{wall} = WT_{init} + WT_{map} + WT_{shuffle} + WT_{reduce} + WT_{end} \quad (1)$$

where $WT_{init}, WT_{end}$ are constant, which denote MapReduce job initialization and conclusion time.

$WT_{map}$ is wall time in Map phase that is presented in equation (2). It depends on the longest map task, namely the worst case a Map phase is decided by task(s) contain all large size of images.

$$WT_{map} = WT_{mapIn} + WT_{mapProcess}$$
$$= discR(SizeBig \cdot \eta) + bdw(SizeBig \cdot \eta) + discW(SizeBig \cdot \eta) + avgANTS(\eta) \quad (2)$$

The worst-case time for shuffle phase depends on all unbuffered data that needs to be retrieved from local disc, transfer through network and then transfer through band as shown in equation (3).. $WT_{reduce}$ is wall time in Reduce phase as equation (4) demonstrated.

$$WT_{shuffle} = discR(SizeGen) + bdw\left( \alpha \cdot \left\lfloor \frac{\#img}{\eta} \right\rfloor \cdot SizeGen \right) + discW\left( \left\lfloor \frac{\#img}{\eta} \right\rfloor \cdot SizeGen \right) \quad (3)$$

$$WT_{reduce} = WT_{reduceProcess} + WT_{reduceOut} = avgA\left( \left\lfloor \frac{\#img}{\eta} \right\rfloor \right) + discR(SizeGen) + discW(SizeGen) \quad (4)$$

Equation (5) displays resource time, which sums up all the time from map, shuffle and Reduce phase as Figure 2 shows.

$$RT = RT_{map} + RT_{shuffle} + RT_{reduce} \quad (5)$$

Similar with wall time, the worst-case resource time involve all big size image retrieval either through local disc or network, and it sums up all map tasks' images average time presented in equation (6).

$$RT_{map} = RT_{mapIn} + RT_{mapProcess}$$
$$= discR(\#img \cdot SizeBig) + discW(\#img \cdot SizeBig) + bdw\left( \beta \cdot \left\lfloor \frac{\#img}{\eta} \right\rfloor \eta \cdot SizeBig \right) + \left\lfloor \frac{\#img}{\eta} \right\rfloor \cdot avgANTS(\eta) \quad (6)$$

$RT_{shuffle}$ is the resource shuffle time shown in equation (7), which includes all unbuffered data loading via local disc, all data transfer through network and all data writing to temporarily place to local disc in the end of shuffle. $RT_{reduce}$ is wall time in Reduce phase as equation (8) demonstrated that is identical with $WT_{reduce}$.

$$RT_{shuffle} = \alpha \left\lfloor \frac{\#img}{\eta} \right\rfloor \cdot [discW(SizeGen) + discR(SizeGen)] + b$$
$$dw\left( \left\lfloor \frac{\#img}{\eta} \right\rfloor \cdot SizeGen \right) + discW\left( \left\lfloor \frac{\#img}{\eta} \right\rfloor \cdot SizeGen \right) \quad (7)$$

$$RT_{reduce} = RT_{reduceProcess} + RT_{reduceOut} = avgANTS\left( \left\lfloor \frac{\#img}{\eta} \right\rfloor \right) + discR(SizeGen) + discW(SizeGen) \quad (8)$$

### 2.3 NoSQL fast query – new table design scheme

In this subsection, we will discuss how to create age-specific template in HadoopBase-MIP. As potential solution that is presented in Figure 1(C), we need to separate image data with relative indexes (i.e., age, sex, demographics, etc.) into different column families. It is to be noted that if they are in same column family using different column qualifiers, the query can still not avoid image traversal. If users only have image files are needed to store and process, in this case, we recommend users using file path/name as table rowkeys, and same rowkey corresponds to one column family that stores image data, and another column family for storing file size as index. Each file size record occupies only several bytes and can help hierarchical region split policy decides the right split point [2, 17].

This table scheme is simple but useful especially for medical imaging data query to avoid image traversal. Here are several benefits for HadoopBase-MIP interface using our proposed table scheme. First, the separate index column family can help speed up checking if file exists when uploading, retrieving and deleting data. Second, it can also collect group or individual data and skip unnecessary record without traversing image data. Third, summary statistical analysis based on subset of data, i.e., average all female's brain within 20-40 years old, and we all image data in one column, and stores indexes in a separate column. Both columns are in different column family. When lauching map tasks, our target is column 2 so that we can quickly filter which rowkeys are needed. Column 1 shares exactly same rowkey with Column 2, thus we can do data retrieval within the map and keep the data locality.

### 2.4 Experiment Design

Experiment cluster setups for both Hadoop and SGE are presented in Figure 4. Each job takes one CPU core with 4 GB memory available. We estimate achievable empirical average bandwidth as 70 Mb/second; disk read speed as 100 Mb/second with write speed as 65 Mb/second. The metrics to verify our proposed methods are wall clock time and resource time. SGE is empirically used as a baseline comparison.

#### 2.4.1. Datasets

The experiment uses 5,153 T1 images retrieved from normal healthy subjects gathered from [25].

#### 2.4.2 Use case 1: Heterogenous cluster

In [18], we found that Hadoop would spent more wall time than SGE if data is balanced allocated on heterogeneous machines. A new HadoopBase-MIP load balancer result is shown at Table 2 based on number and the performance of the CPU per machine. Our goal is to empirically verify how load balancer can improve the performance of Hadoop in a heterogeneous cluster. We use the same experimental design strategy as [18], which is compressing 5,153 T1 images to the .gz format. Each job compresses only one NiFTI image with 2GB memory available and generate one compressed image. The total input size of the images is 77.4 GB and the processing generates 45.7 GB of compressed files as output. To explore the impacts of processing time, we artificially increase the processing time by adding a sleep function without any data retrieval to make the job length of the experiment take an additional 15 – 105 seconds (10 s, 25 s, 40 s, 55 s, 70 s, 85 s, 100 s, 115 s respectively) on a fixed dataset to mimic different job processing requirements. Each machine was used as a Hadoop Datanode and HBase RegionServer for data locality [26]. All machines were also configured using SGE. An additional machine for both approaches serves as a cluster master.

#### 2.4.3 Use case 2 - Large dataset analysis

We first used NiftyReg [27] to perform rigid affine transformation on all images to register to MNI-305 space template [28, 29]. Our goal is to average all 5,153 datasets using ANTS AverageImages tool. Empirically, the largest file size in the dataset is SizeBig = 20 MB, the smallest file size is SizeSmall = 6 MB, and average generated files size is SizeGen = 21 MB. Based on cluster's configuration, we assessed the range of map chunk size of images as η ∈ [30,160], so we manually increase the chunk size by 5 from 30, and empirically test both Hadoop and SGE scenarios.

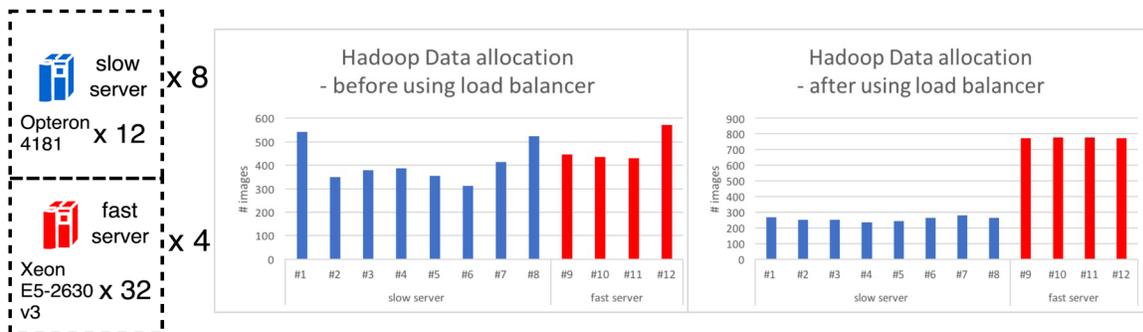

Figure 3. Experiment cluster setup and Data allocation for HadoopBase-MIP. Two different systems (eight machines with 12 slower cores and four machines with 32 fast cores) are used in cluster before applying the load balancer, each machine contains similar amount of image data. After using the load balancer, the data allocations match the ratio #CPU*MIPS.

### 2.4.4 Use case 3 – Rapid NoSQL query

This use case aims to verify the benefit of using proposed table design scheme. Two major population based study features that we are concerned to do average are: age and sex. There are 10 experiments whose setup can be seen in Table 3. Two Hadoop approaches are deigned and compared. A naïve table scheme is to store data of all images, index, sex and age information in a same column family. Our proposed one is to save image data in a separate column family, while the reset index and population info is in another column family. Two Hadoop scenarios are compared with baseline SGE performance. We empirically set 50 images per chunk for one map task.

Table 3. Hadoop v.s. SGE experiment cluster setup with same memory allocation and fixed datasets.

| Experiment | 1 | 2 | 3 | 4 | 5 | 6 | 7 | 8 | 9 | 10 |
|---|---|---|---|---|---|---|---|---|---|---|
| Age(years) | All | | 4-20 | | 20-40 | | 40-60 | | >60 | |
| Female(#people) | 2370 | | 1157 | | 651 | | 230 | | 332 | |
| Male(#people) | | 2120 | | 698 | | 648 | | 280 | | 494 |

### 3. RESULTS

**Use case 1: Heterogeneous cluster**

Figure 3 presents the verification on both modeled and empirical result for Hadoop (before / after using load balancer) and SGE. The wall-clock time performance in Figure 4(A) for SGE is initially limited because of network saturation. As the processing time of a single job increases, SGE performance is linearly increases with lessened network saturation. For both Hadoop scenarios, initial overhead is high when job processing time is small. We can see that as the job processing time increases, Hadoop without load balancing performs worse than SGE, while Hadoop with load balancer performs better than SGE with similar trend. Figure 4(B) shows an aggressive upper bound of theoretical upper bound for SGE; both Hadoop scenarios spend less resource time than SGE, and Hadoop with load balancer performs a little bit better when processing time is less than 40s.

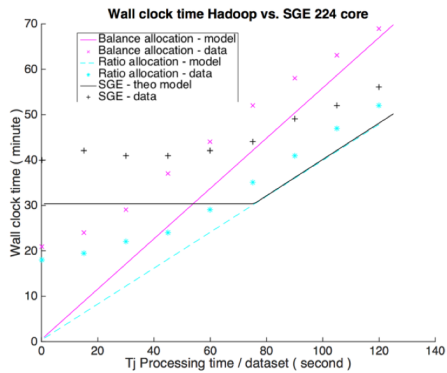

Figure 3 (A) Wall-clock time performance for Hadoop and SGE with different data allocation

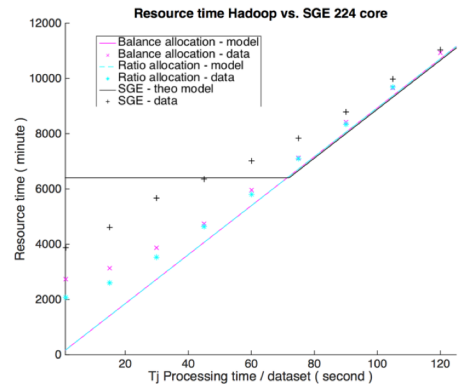

Figure 3 (B) Resource time performance for Hadoop and SGE with different data allocation

**Use case 2: Large datasets analysis**

Figure 4 presents the validation of Hadoop and SGE on averaging 5,153 T1 images. SGE spends 5-fold of wall time and up to 20-fold of resource time more than Hadoop that are presented at Figure 4 (A, B). Figure 5 (C, D) shows the empirical wall time model has same trend with its theoretical model. When the map task chunk size is 50-60, we can observe an optimized wall time. The Hadoop resource model shows an aggressive upper bound, and reveals when chunk size increases to more than 80, resource time are similar. The empirical Hadoop result shows when chunk size more than 80, the resource time becomes similar. The qualitative difference between 5,153 images average and MNI-305 space is shown in Figure 5.

## 3.3 Use case 3: Rapid NoSQL query

When averaging large subsets like all female and all male's T1 images, SGE spends about 3-fold wall time and 6-fold resource time more than proposed Hadoop time. As the size of subsets decreases, we can see that SGE's wall / resource time also decreases, and proposed Hadoop table scheme design also generates similar decreasing trend, and use less wall / resource time than SGE. However, naïve table design scheme leads to opposite trend, for example, averaging all male's T1 images within 40-60 years old cost naïve Hadoop scenario namely when subset's size is small, it spends 6.5-fold wall time more than SGE and about 9-fold more than proposed Hadoop, and 7-fold resource time more than SGE with 12-fold more than proposed Hadoop.

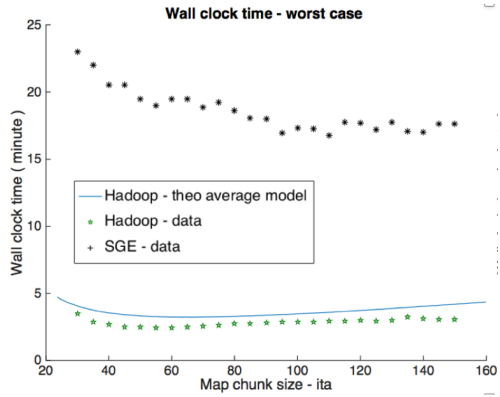

Figure 4 (A) Wall-clock time performance for Hadoop and SGE on large datasets analysis

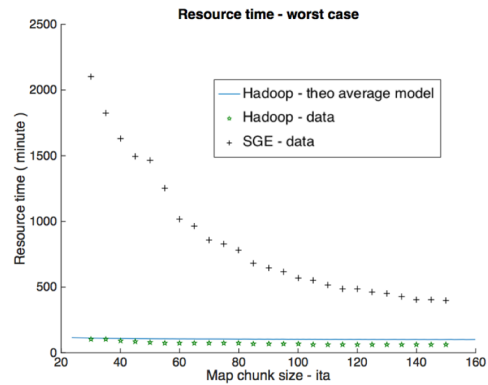

Figure 4 (B) Resource time performance for Hadoop and SGE on large datasets analysis

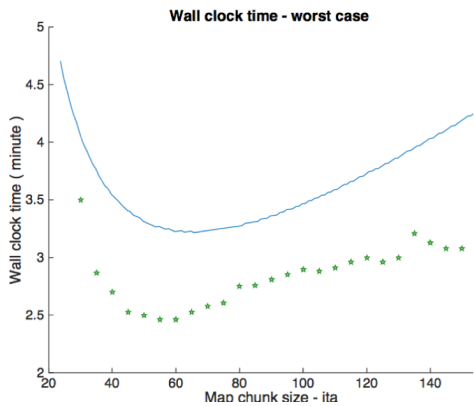

Figure 4 (C) Wall-clock time performance for Hadoop and theoretical model

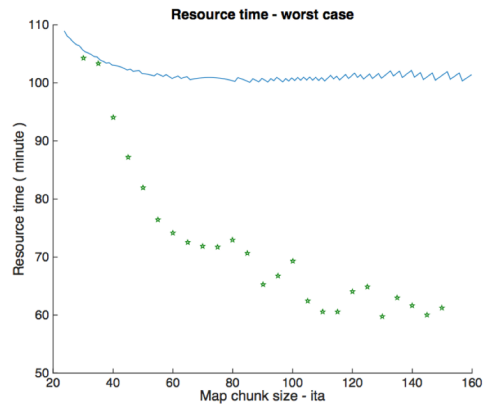

Figure 4 (D) Resource time performance for Hadoop and theoretical model

## 4. CONCLUSION AND DISCUSSION

The paper presents HadoopBase-MIP backend design which is a data Colocation Grid Framework for Big Data MIP. We use three use cases to validate the performance and the usage of the system interface, all cases are compared with SGE. For heterogamous cluster case, SGE's trend reveals that its performance will be degraded by network saturation. When network saturation's impact decreases, Hadoop without load balancer performs worse than SGE due to low utilization of cluster CPU resource. Hadoop with load performance can help reduce this resource overload / starvation issue. The reason of both Hadoop scenarios wall time is high when job processing time is very small reveals when data processing type is high read / write type, the write performance would make congestion of the whole process.

For large dataset analysis case, image averaging analysis is a read intensive with short computation processing. SGE's performance stuck due to data movement via network. Combining theoretical wall time with resource time model for Hadoop summary statistic, we can conclude 50-60 optimal map task chunk. For rapid NoSQL query case, SGE's performance is correlated with the subsets size that needs to do averaged, since there is no query efficiency issue, all data has to be retrieved from network storage to cluster computation node. For naïve Hadoop table scheme, it costs more time than proposed Hadoop and SGE especially when subsets size is relatively small. The reason for this is this scheme would force query traverse image data that is not needed which takes more time than proposed Hadoop scheme that skip small index without image traversal. In summary, HadoopBase-MIP system provides a complete backend interface developed upon built in Hadoop and HBase API.

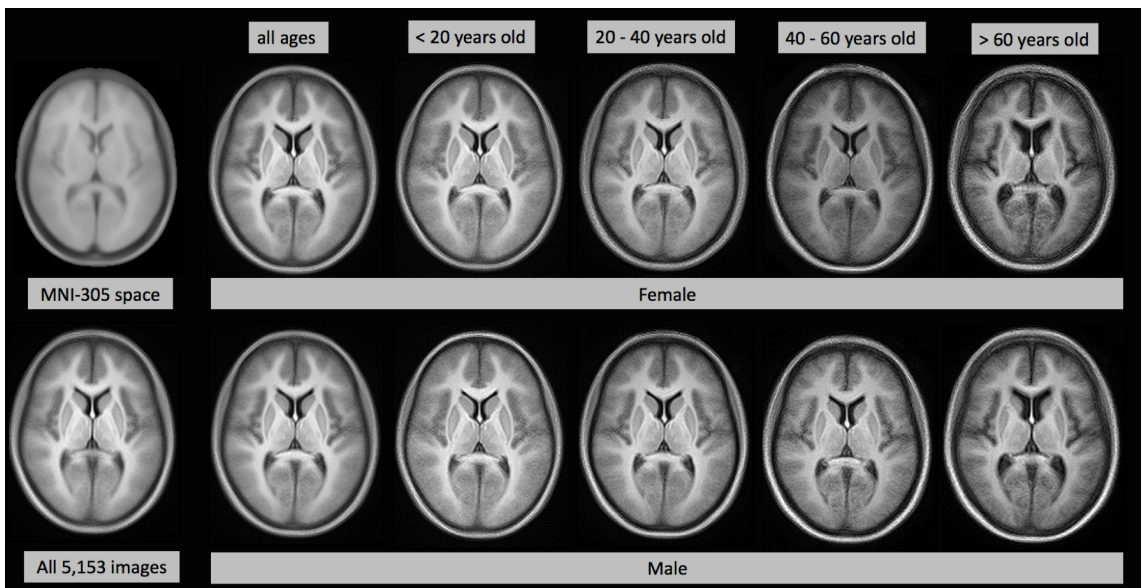

Figure 5. Qualitative results for summary statistics analysis on large datasets and age / sex-specific image averaging analysis.

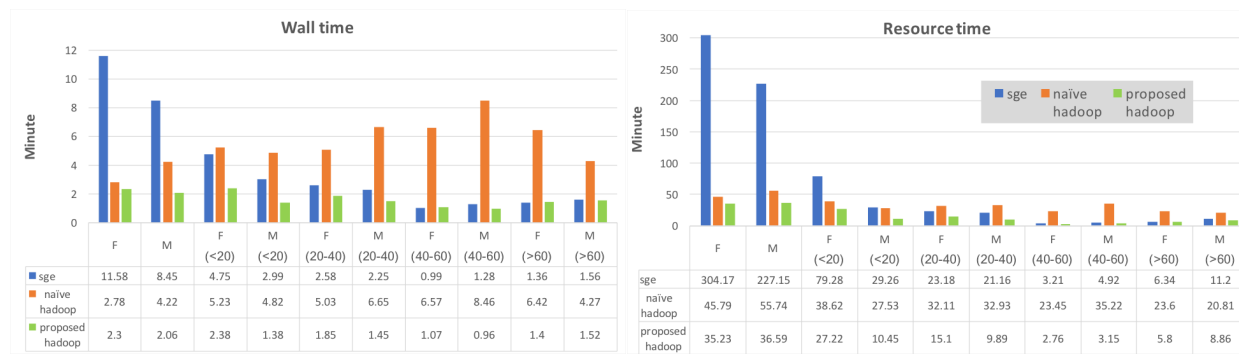

Figure-6 Proposed table scheme design vs. naïve scheme vs. SGE

## ACKNOWLEDGEMENTS

This work was funded in part by NSF CAREER IIS 1452485 and in part by the National Institute on Aging Intramural Program. We are appreciative of the many anonymous volunteers who contributed their time and data to create the de-identified imaging resources used in this study. Any opinions, findings, and conclusions or recommendations expressed in this material are those of the author(s) and do not necessarily reflect the views of NSF. This study was in part using the resources of the Advanced Computing Center for Research and Education (ACCRE) at Vanderbilt University, Nashville, TN. This project was supported in part by ViSE/VICTR VR3029 and the National Center for Research Resources, Grant UL1 RR024975-01, and is now at the National Center for Advancing Translational Sciences, Grant 2 UL1 TR000445-06.